\documentclass{PoS}
\pdfoutput=1

\usepackage[utf8]{inputenc}
\usepackage{amsmath}
\usepackage{amsfonts}
\usepackage{amssymb}
\usepackage{graphicx}
\usepackage{enumitem}

\DeclareMathOperator\Trace{tr}
\DeclareMathOperator\tr{tr}
\newcommand{\Nf}{{N_f}}
\newcommand{\Sdet}{\sigma}
\newcommand{\tphi}{\psi}
\newcommand{\Ns}{N_s}
\newcommand{\D}[2]{D_{#1,#2}}
\newcommand{\Y}{O}

\title{Evading the sign problem in random matrix simulations}

\ShortTitle{Evading the sign problem in random matrix simulations}

\author{\speaker{Jacques C.R. Bloch}%
        \thanks{Supported by the DFG collaborative research center SFB/TR-55 "Hadron Physics from Lattice QCD".}\\
       University of Regensburg\\
       E-mail: \email{jacques.bloch@physik.uni-regensburg.de}}

\abstract{In this talk we show how the sign problem, occurring in dynamical simulations of random matrices at nonzero chemical potential, can be avoided by judiciously combining matrices into subsets. One can prove that these subsets have real and positive weights such that importance sampling can be used in Monte Carlo simulations. The number of matrices per subset is proportional to the matrix dimension. We measure the chiral condensate and observe that the statistical error is independent of the chemical potential and grows linearly with the matrix dimension, which contrasts strongly with its exponential growth in reweighting methods.}

\FullConference{The XXIX International Symposium on Lattice Field Theory - Lattice 2011\\
		 July 10-16, 2011\\
		 Squaw Valley, Lake Tahoe, California}

\begin{document}

\section{Introduction}

Dynamical Monte Carlo simulations of QCD are seriously hampered at nonzero chemical potential $\mu$ because the fermion determinant becomes complex, causing the notorious sign problem \cite{deForcrand:2010ys}.
The sign problem in QCD can be explored using random matrix theory (RMT) \cite{Splittorff:2007ck} because of the equivalence between QCD in the $\epsilon$-regime and RMT \cite{Basile:2007ki}.
Although many observables in unquenched RMT have been computed analytically, it is interesting to investigate if one could also access these numerically.
Dynamical simulations of random matrices at nonzero ${\mu}$ also suffer from a sign problem and can therefore be used as a playground for algorithmic developments.
In this talk we will present a subset method, which solves the sign problem in the RMT case by judiciously combining matrices into subsets with real and positive fermionic weights \cite{Bloch:2011jx}.

\section{Fermion determinant and sign problem in QCD}

After integration over the fermion fields the QCD partition function can be written as:
\begin{align}
Z_\text{QCD} &=  \int {\cal D} A_\mu \, \underbrace{ e^{- S_G} {\prod_{f=1}^\Nf \det[D_{\mu,m_f}]}  }_{\text{MCMC weight function P ?}} ,
\end{align}
where only the integration over the gauge fields $A_\mu$ remains, $S_G$ is the gauge action and $D_{\mu,m_f}$ is the Dirac operator for a quark of mass $m_f$ at chemical potential $\mu$.
As long as the weight factors are real and positive, the functional integral in lattice QCD is evaluated by a Markov chain Monte-Carlo (MCMC) simulation using importance sampling, and the expectation value of an observable $\Y$ is approximated by the sample average of the measurements on $N_\text{MC}$ configurations:
\begin{align}
\overline \Y = \frac{1}{N_{\text{MC}}} \sum_{j=1}^{N_\text{MC}}  \Y_j .
\label{sampavg}
\end{align}

The fermion determinant $\det[D_{\mu,m}]$ is real and positive for $\mu=0$, but becomes complex when $\mu \neq 0$. In the latter case the fermion determinant can no longer be interpreted as a probabilistic weight in MCMC simulations and we are confronted with the sign problem.
Methods to perform measurements at finite chemical potential, by circumventing the sign problem, generically require a computing time which grows  exponentially with the volume. This is, for example, the case in reweighting methods, where the ensemble is sampled according to an auxiliary weight function and the results are reweighted appropriately.

\section{Random matrix theory}

\begin{figure}
\centerline{\includegraphics[width=0.8\textwidth,clip]{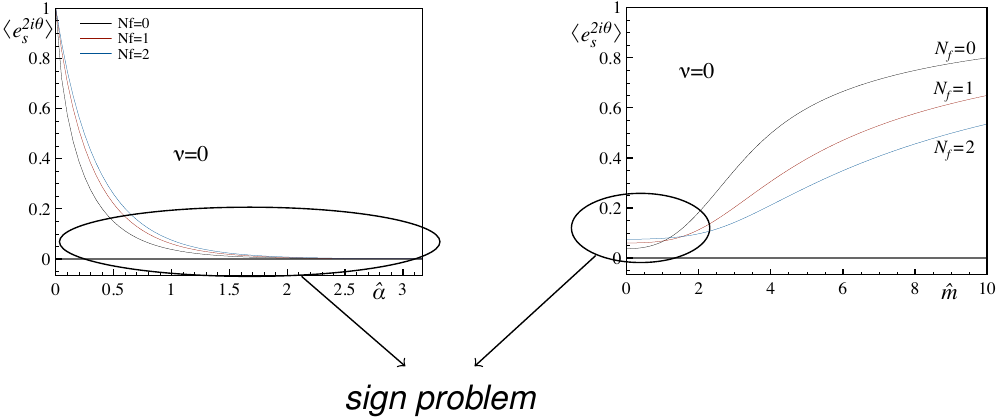}}
\caption{Average phase factor $\langle e^{2 i \theta}\rangle$ as a function of the chemical potential $\hat\alpha=2N\mu^2$ for $m=0$ (left) and as a function of the mass $\hat m= 2N m$ for $\hat\alpha=1$ (right). A small average phase factor corresponds to a strongly fluctuating phase and is evidence of the sign problem.}
\label{avgphase}
\end{figure}

In the $\varepsilon$-regime, QCD is equivalent to chiral random matrix theory, both at zero and nonzero chemical potential \cite{Basile:2007ki}.
In the two-matrix model of Osborn \cite{Osborn:2004rf} the random matrices $\phi_1$ and $\phi_2$ are complex $(N+\nu) \times N$ matrices distributed according to the unquenched partition function
\begin{align}
Z_\nu^{\Nf}(\mu;\{m_f\})  = \int d\phi_1 d\phi_2 \, w(\phi_1) \, w(\phi_2) \, {\prod_{f=1}^\Nf \det D_{\mu,m_f}(\phi_1, \phi_2)}
\label{Zrmt}
\end{align}
with Gaussian weights
$w(\phi) = (N/\pi)^{N(N+\nu)} \exp(-N \Trace \phi^\dagger \phi)$ and $N_f$ dynamical quarks of masses $m_f$ at a chemical potential $\mu$, whose Dirac operator (with $\nu$ zero modes) is given by:
\begin{align}
D_{\mu,m}(\phi_1,\phi_2) = 
\begin{pmatrix}
m & i\phi_1 + \mu \phi_2 \\
i \phi_1^\dagger + \mu \phi_2^\dagger & m
\end{pmatrix}.
\label{model}
\end{align}
The dynamics of the random matrix model crucially depend on the determinant of the Dirac operator.
Just as in QCD, the Dirac matrix $D_{\mu,m}$ is non-Hermitian in this random matrix model for ${\mu} \neq 0$: Its determinant is complex and can be written as $\det[D_{\mu,m}]  \equiv R e^{i{\theta}}$.
The average phase factor $\langle e^{2i\theta} \rangle$ reflects the fluctuations of the fermion determinant and characterizes the strength of the sign problem in dynamical simulations. It was computed analytically in refs.~\cite{Splittorff:2007ck,Bloch:2008cf,Bloch:2011jk} and is shown in fig.~\ref{avgphase}, where we highlighted the parameter region where the sign problem occurs.

In this paper we examine the sign problem in dynamical simulations of chiral random matrices. Before presenting a new solution for the sign problem we briefly describe the standard reweighting method, which we will later use to compare with the new results and to verify the onset of the sign problem.

\section{Reweighting}

The reweighting method can be used to circumvent the sign problem and perform measurements in dynamical simulations with complex weights.
The ensemble average of an observable $y(x)$ in an ensemble with weight $w(x)$ is defined by
\begin{align}
\langle {y} \rangle_{w} = \frac{\int dx \; w(x) y(x)}{\int dx \; w(x)} .
\end{align}

In the reweighting method one introduces an auxiliary ensemble with weight $w_\text{aux}(x)$ and rewrites the previous equation as
\begin{align}
\langle {y} \rangle_{w} = \frac{\int dx \; w_\text{aux}(x) \frac{w(x)}{w_\text{aux}(x)} y(x)}{\int dx \; w_\text{aux}(x) \frac{w(x)}{w_\text{aux}(x)}} 
= \frac{\left\langle {\frac{w}{w_\text{aux}} y} \right\rangle_{\!w_\text{aux}}}{\left\langle {\frac{w}{w_\text{aux}}} \right\rangle_{\!w_\text{aux}}} .
\label{reweight}
\end{align}
If the ensemble $w_\text{aux}$ is chosen to be real and positive it can be sampled using importance sampling methods and the result of eq.~\eqref{reweight} can be evaluated in a Monte Carlo simulation.
Typical examples for $w_\text{aux}$ are the quenched, phase-quenched, $\mu$-quenched and sign-quenched ensembles.

The problem with reweighting methods is that the work needed to make reliable measurements on the statistical ensemble grows exponentially with volume and chemical potential because it involves the computation of exponentially small reweighting factors from a statistical sampling of largely canceling contributions \cite{deForcrand:2010ys}.

\section{Subset method for dynamical RMT simulations}

Below we describe a subset method which solves the sign problem for dynamical simulations of the Osborn model and was first introduced in ref.~\cite{Bloch:2011jx}.

For any given random matrix pair $\phi=(\phi_1, \phi_2)$ we introduce a set of matrices
\begin{align}
\Omega(\phi)=\left\{ \tphi(\phi;\theta_n): \theta_n = \frac{\pi n}{\Ns} \wedge n=0,\ldots,\Ns\!-\!1 \right\} ,
\label{subset}
\end{align}
containing $N_s$ orthogonal rotations  $\tphi=(\tphi_1, \tphi_2)$ of $\phi$ defined by 
\begin{align}
\begin{pmatrix}
\tphi_1(\phi;\theta) \\ \tphi_2(\phi;\theta) 
\end{pmatrix}
=
\begin{pmatrix}
\;\;\;\;\cos\theta & \sin\theta \\
- \sin\theta & \cos\theta 
\end{pmatrix}
\begin{pmatrix}
\phi_1\\\phi_2
\end{pmatrix} .
\label{ortrot}
\end{align}
The subset construction \eqref{subset} allows for a partial resummation of the original random matrix partition function \eqref{Zrmt}, which can be rewritten as an equivalent partition function over subsets $\Omega$: 
\begin{align}
  Z = \int d\Omega \, W(\Omega) \, \Sdet_\Omega(\mu,m)  .
\label{Zsubset} 
\end{align}
The subset weights were factorized in a Gaussian part $W(\Omega) \equiv w(\tphi_1(\phi;\theta) )w(\tphi_2(\phi;\theta) )$, which is independent of $\theta$ because of the orthogonal rotations \eqref{ortrot}, and a fermionic weight
\begin{align}
\Sdet_\Omega(\mu,m) = \sum_{n=0}^{\Ns-1} {\det}^{N_f} \D{\mu}{m}(\tphi(\phi;\theta_n))   ,
\label{fw}
\end{align}
which is a sum of complex determinants.
The equivalence of the partition functions \eqref{Zrmt} and \eqref{Zsubset} rests on the observation that there is a subset ${\Omega(\phi)}$ for each configuration $\phi=(\phi_1,\phi_2)$ of the original partition function, so that the set of all subsets forms an $\Ns$-fold covering of the original RMT ensemble.

The subset method solves the sign problem because of the following \textit{positivity theorem:} 
For any subset $\Omega$ given by eq.~\eqref{subset} the fermionic subset weight $\Sdet_\Omega(\mu,m)$ is {\textit{real}} and {\textit{positive}} if ${\Ns > N_f N}$ (for arbitrary $m$ and $\mu<1$ ).

This theorem results from the following identity, which relates the fermionic subset weights at nonzero and zero chemical potential. For arbitrary $\mu$ and $m$:
\begin{align}
\Sdet_\Omega(\mu,m)  = (1-\mu^2)^{N_f(N+\frac{\nu}{2})} \Sdet_\Omega\left(0,\frac{m}{\sqrt{1-\mu^2}}\right) ,
\label{posrel}
\end{align}
for any $\Omega$ constructed according to \eqref{subset} \textit{if} $\Ns > N_f N$. The proof of this identity will be given in a forthcoming publication.  From this identity the positivity of $\Sdet_\Omega(\mu,m)$ is easily derived: For $\mu=0$ the determinants of all the matrices in the subset are real and positive, as all the eigenvalues of the Dirac matrix come in complex conjugate pairs in this case (for arbitrary real mass). Therefore, eq.~\eqref{posrel} implies that the fermionic weight $\Sdet_\Omega(\mu,m)$ is real and positive for $\mu<1$. Moreover, for $\mu=1$ and $m=0$ the sum of determinants is exactly zero, which corresponds to the case of maximal non-hermiticity.

Note that eq.~\eqref{posrel} is an extension to arbitrary mass of the identity originally given in ref.~\cite{Bloch:2011jx}, which only covered the massless case, while an inequality described the case $m \neq 0$.

\section{Simulations}

The positive subset weights $W(\Omega)\Sdet_\Omega(\mu,m)$ were used to generate subsets of random matrices and sample the partition function \eqref{Zsubset} with a Metropolis algorithm. In practice the subset size is set to $\Ns=N_f N+1$, which is the minimum value for which the positivity of the fermionic weights is guaranteed.
Successive subsets in the Markov chain are generated as follows: 
\begin{itemize}[itemsep=-0.5mm,topsep=1mm]
\item randomly choose a configuration in the current subset,
\item generate a new configuration by making a random step, 
\item construct the subset corresponding to the new configuration, 
\item apply the accept-reject step to the proposed subset using the positive subset weights. 
\end{itemize}
This algorithm satisfies detailed balance and ergodicity such that the partition function will be sampled correctly by the MCMC algorithm.
In the subset method the sample average $\overline \Y$ measured over a sample of $N_\text{MC}$ subsets $\Omega_k$, approximating the ensemble average in the original RMT ensemble, is computed by
\begin{align}
  \overline \Y_{\mu,m} &= \frac{1}{N_\text{MC}} \sum_{k=1}^{N_\text{MC}}
\sum_{n=0}^{\Ns-1} \frac{{\det}^{N_f} \D{\mu}{m}(\psi^{kn})}{\Sdet_{\Omega_k}(\mu,m)} \, \Y_{\mu,m}(\psi^{kn}) ,
\label{observ}
\end{align}
where $\psi^{kn} \in\Omega_k$ and one takes into account that the matrices inside the subsets yield different values for the measured observable.

\section{Results}

We applied the subset method to compute the chiral condensate 
\begin{align*}
\Sigma=\frac{1}{2N} \frac{1}{Z} \frac{dZ}{dm} = \left\langle \frac{1}{2N} \tr \det D^{-1}_{\mu,m}(\phi_1, \phi_2) \right\rangle
\end{align*}
in the RMT model.
In each Markov chain we generated $N_\text{MC}$=100,000 subsets. 
The subsets in the Markov chain are correlated, producing $N_\text{MC}/2\tau$ independent measurements for an integrated autocorrelation time $\tau$.
The statistical errors are determined taking the autocorrelations into account.

The results of the subset method are compared with those computed with standard reweighting methods. For the latter we generate $N_\text{MC} \times \Ns$ random matrices, such that the total number of matrices is the same as in the subset method. 
 
\begin{figure}
\centerline{%
\includegraphics[width=0.33\textwidth]{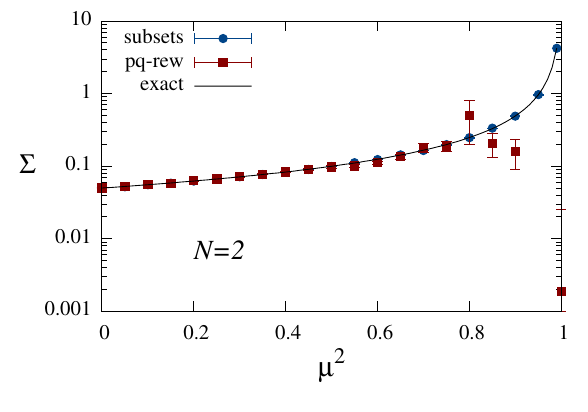}%
\includegraphics[width=0.33\textwidth]{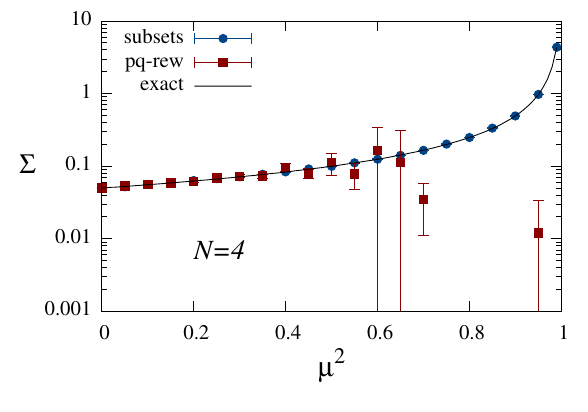}%
\includegraphics[width=0.33\textwidth]{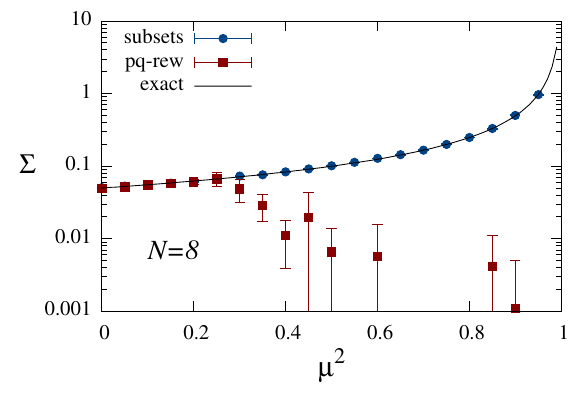}}

\centerline{%
\includegraphics[width=0.33\textwidth]{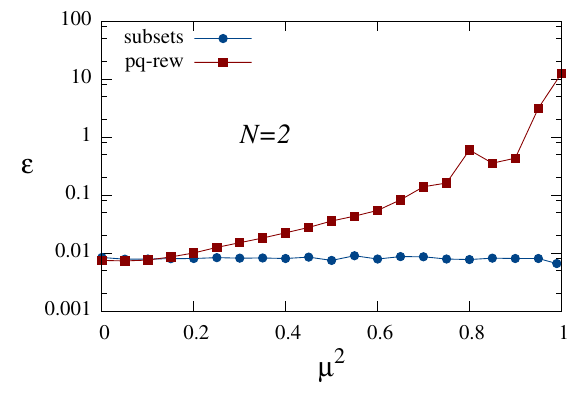}%
\includegraphics[width=0.33\textwidth]{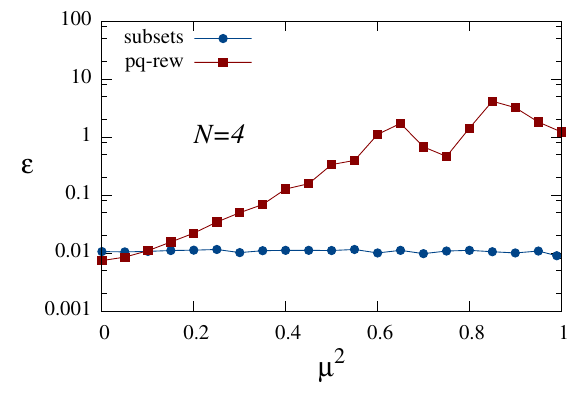}%
\includegraphics[width=0.33\textwidth]{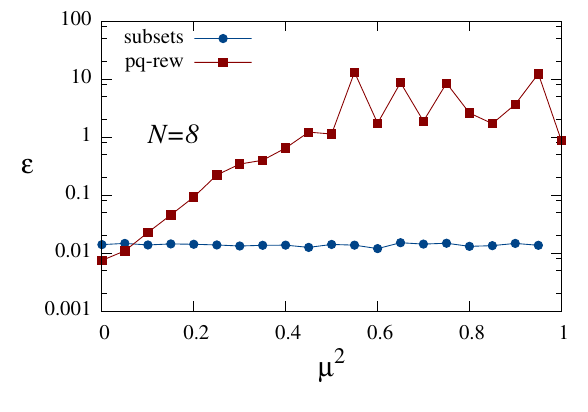}}
\caption{Top row: chiral condensate $\Sigma$ versus chemical potential $\mu^2$ for the subset method (blue bullets) and phase-quenched reweighting (red squares) for $N=2,4,8$.
The full line corresponds to the exact analytical result \cite{Osborn:2008jp}. Bottom row: relative statistical error $\varepsilon$ for the same data. The error for the reweighting method grows very rapidly and can only be trusted as long as the overlap problem is under control.
}
\label{cc_vs_mu2}
\end{figure}

Simulations were performed for $N=2,\ldots,34$ with $N_f=1$ and $m=0.1/2N$ (the mass is small w.r.t. the magnitude of the smallest eigenvalue).
In fig.~\ref{cc_vs_mu2} the results for the chiral condensate $\Sigma$ (top row) and its relative statistical error $\varepsilon$ (bottom row) are shown as a function of the chemical potential. 
The statistical error of the phase-quenched reweighting grows exponentially with $\mu$, until the method fails when the set of sampled matrices no longer overlaps with the relevant configurations. As the matrix size increases this failure occurs for smaller and smaller $\mu^2$. This strongly contrasts with the subset method where the results are reliable up to much larger values of $\mu^2$ and agree with the analytical predictions of ref.~\cite{Osborn:2008jp}. Moreover, the error is independent of the chemical potential.
\begin{figure}[b]
\centerline{\includegraphics[width=0.4\textwidth]{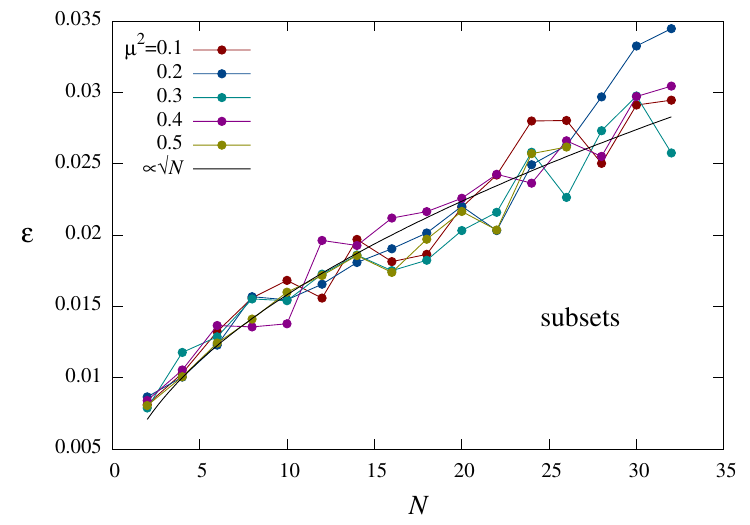}\hspace{5mm}
\includegraphics[width=0.4\textwidth]{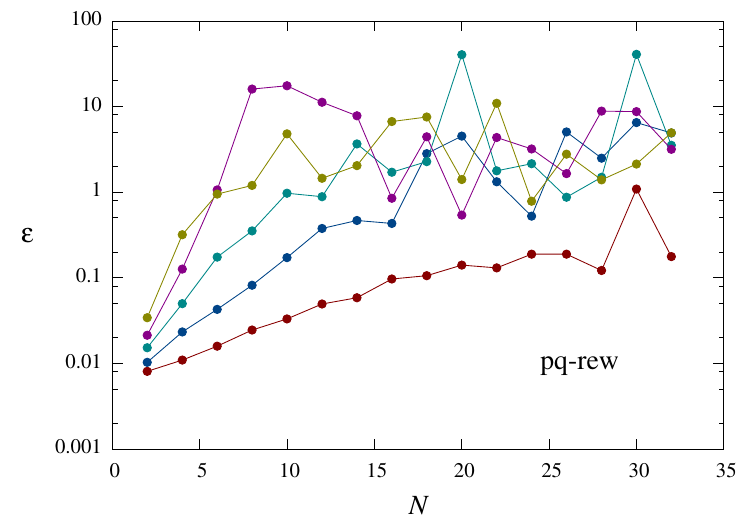}}
\caption{
Relative error $\varepsilon$ on the chiral condensate versus matrix size $N$ for various values of the chemical potential. The results for the subset method are given on the left, those for phase-quenched reweighting on the right. Note that the later are given on a semi-log plot due to the exponential increase of the error.
}
\label{cc_vs_N}
\end{figure}
We also investigated how the statistical error depends on the matrix size, see fig.~\ref{cc_vs_N}. As expected, the work grows exponentially with $N$ for the reweighting method (right). However, for the subset method (left) the error is proportional to $\sqrt{N}$ for a fixed number of sampled subsets, or $N$ for a fixed number of sampled matrices. Conversely, to achieve a constant error the number of sampled matrices should grow as $N^2$. 

A comparison of both methods makes clear how the sign problem is solved in the subset method: The cancellations needed to yield the exponentially small numbers in the partition function no longer happen through statistical sampling of the ensemble, but occur deterministically inside subsets of size of ${\cal O}(N)$. Therefore, no exponential increase of the computing time with volume and chemical potential is expected in the subset method, as was confirmed by the numerical results. 

Note that in the reported study the fermionic weights \eqref{fw} were directly computed by summing over the complex determinants at chemical potential $\mu$. However, we could equally well use eq.~\eqref{posrel} to compute the exponentially small subset weights at $\mu$ from those at $\mu=0$, hence avoiding the need for any numerical cancellations. This strategy would be accompanied by some overhead because the determinants have to be computed both at $\mu=0$, to compute the subset weights, and at $\mu$ to compute the observable using eq.~\eqref{observ}.

\section{Summary}
 
In this talk I have discussed the sign problem occurring in dynamical simulations of random matrices and presented a subset method which solves this problem for the Osborn model. The main feature of the method is the construction of subsets of matrices for which the fermionic weights, i.e. the sums of complex fermion determinants, are real and positive. Importance sampling methods can be used to construct Markov chains of subsets and sample the random matrix ensemble. The numerical simulations confirmed that the subset method solves the sign problem for this model.

The method relies on the ability to construct subsets with positive weights, and an important question arises as to what conditions need to be satisfied to apply this method to relevant physical systems. This matter will be explored further in future research.

\bibliographystyle{jbJHEP}
\bibliography{biblio}

\end{document}